# Hybrid Monte Carlo Simulation of Polymer Chains

*A. Irbäck*

Department of Theoretical Physics, University of Lund
Sölvegatan 14A, S-223 62 Lund, Sweden

## Abstract

We develop the hybrid Monte Carlo method for simulations of single off-lattice polymer chains. We discuss implementation and choice of simulation parameters in some detail. The performance of the algorithm is tested on models for homopolymers with short- or long-range self-repulsion, using chains with $16 \leq N \leq 512$ monomers. Without excessive fine tuning, we find that the computational cost grows as $N^{2+z'}$ with $0.64 < z' < 0.84$. In addition, we report results for the scaling of the end-to-end distance, $r_{1N} \sim N^\nu (\ln N)^{-\alpha}$.

# 1 INTRODUCTION

Monte Carlo methods is a well-established tool in the study of polymer models and many efficient algorithms have been developed [1]. To facilitate the study of large systems it is, nevertheless, of interest to continue the search for improved methods. In this paper we investigate the hybrid Monte Carlo method (HMC) [2], which can be used to simulate off-lattice chains. Although based on molecular dynamics, the method has, to our knowledge, not been applied to polymer models before.

In HMC the system is evolved by using molecular dynamics in a fictitous time, which is supplemented with refreshments of the "momenta". The algorithm is made exact by incorporating a Metropolis type accept-reject step, which eliminates finite-step-size errors arising from the discretization of the equations of motion. The method has become widely used for simulations of lattice QCD with dynamical fermions. Important there is that a HMC update of the whole system requires only O(1) calculations of the highly non-local Boltzmann weight. In this paper we apply HMC to single chains with interactions between all monomer pairs. A whole chain of this kind can be updated in a computer time of order $N^2$. As an example, let us compare this with the behaviour of the popular pivot algorithm [3]. This algorithm was thoroughly tested for self-avoiding walks on a lattice in ref. [4], and was found very powerful for generating independent measurements of global quantities. However, the computer time required to generate one independent measuremnet of local quantities grows as $N^3$ or faster for the chains considered here, since each elementary pivot move scales as $N^2$. The hope is that HMC can be used to improve on this, without losing good efficiency for global quantities.

There are different ways to implement HMC, and the resulting efficiency varies. In our calculations we employ the so-called Fourier acceleration technique [5]. For linear chains with free endpoints, this is equivalent to performing ordinary HMC updates of the bond variables. We have tested this algorithm on four models with self-repulsion. We find that the computational cost of generating one independent measurement grows as $N^{2+z'}$ with $z'$ beween 0.64 and 0.84, and that it is similar for global and local quantities. In particular, these results suggest that thermalization can be carried out faster with HMC than with the pivot algorithm for large $N$. This property is important since thermalization can be the dominating cost with the pivot algorithm.

The plan of this paper is as follows. In section 2 we present the models considered. In section 3 we describe the algorithm and the tuning of simulation parameters. The measurements of the efficiency of the algorithm are discussed in section 4, where we also present results for the scaling of the end-to-end distance. A summary is given in section 5.

# 2 THE MODELS

Our tests of the performance of HMC have been carried out on models with long- or short-range self-repulsion. We first consider a polyelectrolyte in a solution. Here the repulsive interaction is taken to be of screened Coulomb type, and the presence of a salt concentration is parametrized through a finite Debye screening length. The full potential energy considered

is

$$E = \frac{k}{2} \sum_{i=1}^{N-1} r_{i\,i+1}^2 + \frac{q^2}{4\pi\epsilon_r\epsilon_0} \sum_{1 \leq i < j \leq N} \frac{e^{-\kappa r_{ij}}}{r_{ij}} ,  \quad (1)$$

where $q$ is the monomer charge, $\epsilon_r\epsilon_0$ the dielectric permittivity of the medium and

$$r_{ij} = \left(\sum_{\mu=1}^{3}(x_{i\mu} - x_{j\mu})^2\right)^{1/2} \quad (2)$$

is the distance between monomers $i$ and $j$. The inverse screening length $\kappa$ at salt concentration $c_s$ is given by

$$\kappa = q\left(\frac{2N_A c_s}{\epsilon_r\epsilon_0 k_B T}\right)^{1/2} , \quad (3)$$

where $N_A$ is Avogadro's number and $T$ the temperature. For convenience, all lengths will be measured in units of $r_0 = (\frac{q^2}{4\pi\epsilon_r\epsilon_0 k})^{1/3}$. The partition function then becomes

$$Z = \int \left[\prod_{i=1}^{N}\prod_{\mu=1}^{3} dx_{i\mu}\right] \prod_{\mu=1}^{3} \delta\left(\sum_{i=1}^{N} x_{i\mu}\right) \exp\left[-\frac{1}{\tilde{T}}\left(\frac{1}{2}\sum_{i=1}^{N-1} r_{i\,i+1}^2 + \sum_{1 \leq i < j \leq N} \frac{e^{-\tilde{\kappa} r_{ij}}}{r_{ij}}\right)\right] , \quad (4)$$

where $\tilde{\kappa} = r_0\kappa$ and $\tilde{T} = (kr_0^2)^{-1}k_B T$. The endpoints of the chains are free and the $\delta$ function is needed because of translational invariance. Our calculations have been performed using $c_s = 0$ (unscreened Coulomb potential) or 1M. The other parameters were taken to be $r_0 = 6$Å, $q = e$, $\epsilon = 78.3$ and and $T = 298$K. This corresponds to $\tilde{T} = 0.838$ and to $\tilde{\kappa} = 1.992$ for $c_s = 1$M. These two sets of parameters were chosen to make a comparison possible with the recent results of refs. [6, 7]. These authors investigated the model using a Gaussian variational method, and performed also Monte Carlo calculations for $N$ up to 2048.

The scaling with $N$ of the end-to-end distance defines the critical index $\nu$ through

$$\langle r_{1N} \rangle \sim N^\nu , \qquad N \to \infty . \quad (5)$$

Logarithmic corrections to this asymptotic relation may appear. The unscreened and screened Coulomb models represent different values of $\nu$. The unscreened potential gives rise to rod-like behaviour with $\nu = 1$, while the Flory result $\nu = 3/(2+d) = 0.6$ is approximately valid for the short-range screened potential.

In addition to these two cases, we also consider two potentials of intermediate range. Here the repulsion energy decays as $r^{-\lambda}$ with $\lambda = 2$ and 2.5, respectively, and the partition function is

$$Z = \int \left[\prod_{i=1}^{N}\prod_{\mu=1}^{3} dx_{i\mu}\right] \prod_{\mu=1}^{3} \delta\left(\sum_{i=1}^{N} x_{i\mu}\right) \exp\left[-\frac{1}{\hat{T}}\left(\frac{1}{2}\sum_{i=1}^{N-1} r_{i\,i+1}^2 + \sum_{1 \leq i < j \leq N} \frac{1}{r_{ij}^\lambda}\right)\right] , \quad (6)$$

where we put the dimensionless temperature parameter $\hat{T} = 1$. This model was investigated in ref. [8] for $2 \leq \lambda < 3$. Using a Gaussian variational approach, these authors predicted that $\nu = 2/\lambda$, which differs from the Flory-type result $\nu = 3/(2+\lambda)$. For $\lambda = 2$, they further predicted the appearance of logarithmic corrections to eq. 5 of the form

$$\langle r_{1N} \rangle \sim N^\nu (\ln N)^{-\alpha} \tag{7}$$

with $\nu = 2/\lambda = 1$ and $\alpha = 1/2$. These results were tested numerically in ref. [9], using chains of size $N \leq 120$. The Monte Carlo data gave support to the result $\nu = 2/\lambda$ and to the presence of logarithmic corrections for $\lambda = 2$. However, the precise value of the exponent $\alpha$ could not be determined from these data. In section 4, we will therefore use our results to get an improved estimate of $\alpha$. We will find that the prediction of ref. [8] is in nice agreement with the data.

Finally, we mention the virial theorem which provides a useful check of the Monte Carlo procedure, as discussed in refs. [6, 7]. For the partition functions in eqs. 4 and 6 this exact identity takes the forms

$$\left\langle \sum_{i=1}^{N-1} r_{i\,i+1}^2 - \sum_{1 \leq i < j \leq N} \frac{e^{-\tilde\kappa r_{ij}}}{r_{ij}} - \tilde\kappa \sum_{1 \leq i < j \leq N} e^{-\tilde\kappa r_{ij}} \right\rangle = 3\tilde T(N-1) \tag{8}$$

and

$$\left\langle \sum_{i=1}^{N-1} r_{i\,i+1}^2 - \lambda \cdot \sum_{1 \leq i < j \leq N} \frac{1}{r_{ij}^\lambda} \right\rangle = 3\hat T(N-1) , \tag{9}$$

respectively.

## 3 THE ALGORITHM

The HMC algorithm [2] is a general method for simulation of statistical systems with continuous degrees of freedom and can be directly applied to the models considered here. Naive HMC updates of the monomer coordinates leads, however, to poor performance due to the slow evolution of long-wavelength modes. For linear chains with free endpoints there are two equivalent ways to alleviate this problem. One is, as in the usual Metropolis algorithm, to instead update the bond variables $b_{i\mu} = x_{i+1\,\mu} - x_{i\mu}$. The other is to employ the Fourier acceleration technique [5].

### 3.1 HMC

We begin with a brief description of ordinary HMC using the bond variables $b_{i\mu}$. To simulate a model defined by the potential energy $E(b)$, one introduces a set of conjugate "momenta" $p_{i\mu}$ and makes use of the auxiliary Hamiltonian

$$H_{MC}(p,b) = \frac{1}{2} \sum_{i\mu} p_{i\mu}^2 + \frac{E(b)}{T} . \tag{10}$$

A finite-step approximation of the equations of motion arising from $H_{MC}$ are used to guide the evolution of the system. A convenient choice of discretization is the leapfrog scheme

$$\begin{cases} b'_{i\mu} &= b_{i\mu} + \epsilon p_{i\mu} - \frac{\epsilon^2}{2}\frac{1}{T}\frac{\partial E(b)}{\partial b_{i\mu}} \\ p'_{i\mu} &= p_{i\mu} - \frac{\epsilon}{2}\frac{1}{T}\left[\frac{\partial E(b)}{\partial b_{i\mu}} + \frac{\partial E(b')}{\partial b_{i\mu}}\right] \end{cases} \qquad i = 1, \ldots, N-1 \quad \mu = 1, 2, 3 \ . \qquad (11)$$

where $\epsilon$ denotes the step size.

The first step in the Monte Carlo procedure is to assign independent random values to the momenta, drawn from the distribution $P(p_{i\mu}) \propto \exp(-\frac{1}{2}p_{i\mu}^2)$. Eq. 11 is then iterated $n$ times, starting from the old configuration $b_{i\mu}$. This is called one trajectory, and we denote the final position of the system by $p^f_{i\mu}, b^f_{i\mu}$. In the last step, the new configuration $b^f_{i\mu}$ is accepted or rejected with probability $\min[1, e^{-\Delta H_{MC}}]$ for acceptance, where $\Delta H_{MC} = H_{MC}(p^f, b^f) - H_{MC}(p, b)$. By using the fact that the phase space map in eq. 11 is reversible and area-preserving, it is easily verified that this final Metropolis step makes detailed balance fulfilled.

The Metropolis question involves the discretization error in the extensive quantity $H_{MC}$. To maintain a reasonable acceptance rate, it is therefore necessary to decrease $\epsilon$ as $N$ is increased. The required variation of $\epsilon$ can be estimated using fairly general arguments [10, 11, 12]. As the number of degrees of freedom tends to infinity, one finds that the mean acceptance becomes

$$P_{acc} = \langle \min(1, e^{-\Delta H_{MC}}) \rangle = \text{erfc}\left(\frac{1}{2}\langle \Delta H_{MC} \rangle^{1/2}\right) , \qquad (12)$$

where $\langle \Delta H_{MC} \rangle \sim N\epsilon^4$ for fixed trajectory length $n\epsilon$. This would imply that constant acceptance requires $\epsilon \sim N^{-1/4}$. Unfortunately, this argument fails for the models considered here. In fact, $\langle \Delta H_{MC} \rangle$, as well as higher moments of $\Delta H_{MC}$, diverges due to the singular nature of the energy function. As the tests below will show, the conclusion remains approximately valid in some of the cases studied, while in other a more rapid decrease of the step size with $N$ is required.

The acceptance rate depends also on the temperature, which so far was assumed to be fixed. This is important to note if the algorithm is used for simulated annealing. If the temperature is lowered, the step size must be decreased to keep the acceptance constant.

The final choice of the two simulation parameters $\epsilon$ and $n$ should be based on autocorrelation properties. The autocorrelation function for a quantity $O$ is given by

$$C_O(m) = \langle O_{m_0+m} O_{m_0} \rangle - \langle O \rangle^2 \qquad (13)$$

where $O_{m_0}$ and $O_{m_0+m}$ are measurements separated by $m$ trajectories. The exponential decay expected at large $m$, $C_O(m) \sim e^{-m/\tau_{exp,O}}$, defines the exponential autocorrelation time $\tau_{exp,O}$. $\tau_{exp,O}$ is the autocorrelation time of the slowest mode that couples to $O$. In our calculations, we consider instead the integrated autocorrelation time

$$\tau_{int,O} = \frac{1}{2}\sum_{m=-\infty}^{\infty} \frac{C_O(m)}{C_O(0)} , \qquad (14)$$

which is usually much easier to measure numerically. $\tau_{int,O}$ directly controls the statistical error on $O$, since the variance of the sample mean is given by

$$\text{var}\Big(\frac{1}{M}\sum_{m=1}^{M} O_m\Big) \approx \frac{C_O(0)}{M} 2\tau_{int,O} \tag{15}$$

for large sample size $M$.

These autocorrelation times are expressed in units of trajectories. To get a measure of computational effort we multiply the autocorrelation time considered, $\tau$, by $n$, since the CPU time required for each trajectory is proportional to $n$. Ideally, the simulation parameters should be chosen so as to minimize the effort $E = n\tau$ for each $N$. This minimization requires extensive testing, which we have not carried out for large $N$. Consequently, we expect that our choice of parameters leads to reasonable but for large $N$ not optimal performance. The corresponding effort is well described by a power law $E \propto N^{z'}$, as we will see below. This means that the required computer time grows as $N^{2+z'}$, since the cost of each step in a trajectory scales as $N^2$.

## 3.2 The Gaussian Chain and Fourier Acceleration

The dynamics of HMC can be analysed in considerable detail for the Gaussian model [13, 14]. We mention here a few results which motivate the Fourier acceleration method. For definiteness, we consider a chain with nearest-neighbour interaction only. We take the potential energy to be

$$\frac{E}{T} = \frac{1}{2}\sum_{i} r_{i\,i+1}^2 = \frac{1}{2}\sum_{i\mu} b_{i\mu}^2 \ . \tag{16}$$

As before, we assume that the center of mass is held fixed and that the endpoints are free. The version of HMC described above is particularly easy to analyse, since $E$ is diagonal in the bond variables. Using this, it is straightforward to compute the exponential autocorrelation time for $b_{i\mu}$ in the limit $\epsilon \to 0$. One finds that $\tau_{exp} = (-\ln|\cos t|)^{-1}$ independent of $i$ and $\mu$, where $t = n\epsilon$ is the trajectory length. Let us now consider HMC in the monomer coordinates, applied to the same model. This case can be analysed by performing the orthogonal transformation

$$\tilde{x}_{k\mu} = \sqrt{\frac{2}{N}} \sum_{i=1}^{N} x_{i\mu} \cos\frac{\pi k(i-\frac{1}{2})}{N} \qquad k = 1,\ldots,N-1 \qquad \mu = 1,2,3 \tag{17}$$

together with the same transformation of the corresponding momenta $\pi_{i\mu}$, which diagonalizes

$$H_{MC} = \frac{1}{2}\sum_{i\mu}\pi_{i\mu}^2 + \frac{1}{2}\sum_{i=1}^{N-1} r_{i\,i+1}^2 = \frac{1}{2}\sum_{k\mu}(\tilde{\pi}_{k\mu}^2 + \omega_k^2 \tilde{x}_{k\mu}^2) \ , \qquad \omega_k^2 = 4\sin^2\frac{\pi k}{2N} \ . \tag{18}$$

In the limit $\epsilon \to 0$, one finds that the exponential autocorrelation time for $\tilde{x}_{k\mu}$ is given by $\tau_{exp,k} = (-\ln|\cos\omega_k t|)^{-1}$. Thus, the long-wavelength modes evolve slowly, as expected. The result above shows this problem can be avoided by using the bond variables. Another way to speed up the evolution of the long-wavelength modes is to increase the step size for these, which is called the Fourier acceleration method. It is clear that this method works for the model in eq. 16 since the different Fourier modes evolve independently. The equations of motion are taken to be

$$\begin{cases} \tilde{x}'_{k\mu} &= \tilde{x}_{k\mu} + \epsilon_k \tilde{\pi}_{k\mu} - \frac{\epsilon_k^2}{2} \frac{1}{T} \frac{\partial E(\tilde{x})}{\partial \tilde{x}_{k\mu}} \\ \tilde{\pi}'_{k\mu} &= \tilde{\pi}_{k\mu} - \frac{\epsilon_k}{2} \frac{1}{T} \left[ \frac{\partial E(\tilde{x})}{\partial \tilde{x}_{k\mu}} + \frac{\partial E(\tilde{x}')}{\partial \tilde{x}_{k\mu}} \right] \end{cases} \qquad k = 1, \ldots, N-1 \quad \mu = 1, 2, 3 \qquad (19)$$

where the varying step size $\epsilon_k = \epsilon/\omega_k$. Except for this change in step size the update is the same as in ordinary HMC.

The Fourier acceleration technique and the use of bond variables are, in general, two distinct ways to improve the efficiency of HMC, and are applicable in different situations. However, they are equivalent for the linear chains with free endpoints considered here. In fact, eq. 11 can be obtained from eq. 19 by performing the same orthogonal transformation of $\omega_k \tilde{x}_{k\mu}$ and $\tilde{\pi}_{k,\mu}$. In the calculations we have used Fourier acceleration since that makes the monomer coordinates more readily available. The cost of the transformation between monomer coordinates and Fourier variables, eq. 17, is negligble since Fast Fourier Transform can be utilized.

The Fourier accelerated algorithm is set up so as to efficiently simulate the particular Gaussian model in eq. 16. The same technique can be used to adapt HMC to general Gaussian models. In the general case, the cost of the required coordinate transformation, corresponding to eq. 17, scales quadratically with $N$. In applications to models with interactions between all the monomer pairs, this is not necessarily severe since the cost of the energy computation already scales as $N^2$. Therefore, a study of this more general class of algorithms could be worthwhile. Especially interesting appears the possibility to make use of the energy function obtained in the Gaussian variational approach. We carried out some tests of this method for the unscreened Coulomb model, using the variational results of refs. [6, 7]. Somewhat disappointingly, however, these tests did not indicate any further gain in efficiency.

The results for the model in eq. 16 suggest one further modification of the algorithm. In fact, the autocorrelation times can get large also with the Fourier accelerated algorithm, if $t$ happens to be near a multiple of $\pi$. To ensure against such accidental mode locking we have, following ref. [15], randomized the trajectory length $t$. Whether this is necessary for the models considered in our applications is not clear. In fact, it is possible that slightly better efficiency could be obtained without this randomization [16].

### 3.3 Tuning of parameters

Next we describe our choice of simulation parameters for the models described in section 2. We begin with the unscreened and screened Coulomb models. Here we have randomized the trajectory length $t = n\epsilon$ by drawing $\epsilon$ from the exponential distribution with mean $\bar{\epsilon}$, while holding $n$ fixed. The average trajectory length $\bar{t} = n\bar{\epsilon}$ has been taken to be independent of $N$, which for the Gaussian model would make the autocorrelation times discussed above independent of $N$. The remaining parameter $\bar{\epsilon}$ has been chosen so as to keep the acceptance constant.

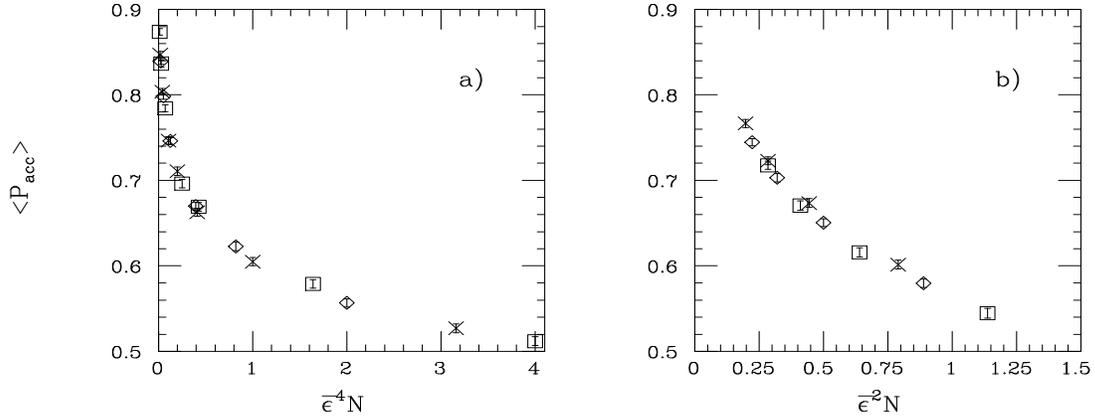

Figure 1: The acceptance probability for a) unscreened and b) screened ($c_s = 1M$) Coulomb potential. The results were obtained for $N = 16$ (crosses), 32 (diamonds) and 64 (squares) and different values of $\bar{\epsilon}$, keeping $\bar{t} = 2$ fixed.

To check the behaviour of the acceptance we performed a set of calculations using $N = 16$, 32 and 64 and various $\bar{\epsilon}$, keeping $\bar{t} = n\bar{\epsilon}$ fixed. Fig. 1a shows the results for the unscreened Coulomb model plotted against $\bar{\epsilon}^4 N$. Approximately, the data fall on a single curve, which would imply that $\bar{\epsilon} \propto N^{-1/4}$ gives constant acceptance. This is the same scaling behaviour as, for example, for the Gaussian model. The results for the screened Coulomb model are shown in fig. 1b and are different. In this case, constant acceptance requires approximately $\bar{\epsilon} \propto N^{-1/2}$. As might have been expected, the Monte Carlo dynamics seems to be more affected by the singularities in the potential when this is short-range.

For the $r^{-\lambda}$ potentials with $\lambda = 2$ and 2.5, we first tried to use the same tuning prescription with $\bar{t}$ independent of $N$. However, the effort $E$ then turned out to increase faster with $N$ than for the previous two models. To improve on this it was necessary to let $\bar{t}$ increase with $N$, and we decided to take $\bar{t} \propto N^{1/2}$. The randomization of $\bar{t}$ has for these two models instead been done, for fixed $\epsilon$, by taking $n$ to be uniformly distributed in $2\bar{n}/3 < n < 4\bar{n}/3$, which improved the efficiency. The acceptance has been kept roughly constant by adjusting $\epsilon$, which can be done by means of short test runs.

For all four models, we carried out a set of test runs to determine approximately optimal parameters for $N = 32$. Starting from these, we have then varied the parameters with $N$ in the way described. We stress that the resulting efficiency is not expected to be optimal for large $N$. Further adjustments of $\bar{t}$ may bring significant improvements.

| $c_s$ | $N$ | Its. | $\bar{t}$ | $n$ | $P_{acc}$ | $\tau_{int,ee}$ | $\lambda$ | $N$ | Its. | $\bar{t}$ | $\bar{n}$ | $P_{acc}$ | $\tau_{int,ee}$ |
|---|---|---|---|---|---|---|---|---|---|---|---|---|---|
| 0 | 16 | 22/2 | 2 | 7 | 0.74 | 2.07(8) | 2 | 16 | 5/1 | 1.6 | 7 | 0.80 | 1.45(11) |
|  | 32 | 22/2 | 2 | 8 | 0.75 | 2.8(2) |  | 32 | 5/1 | 2.2 | 12 | 0.76 | 1.29(12) |
|  | 64 | 22/2 | 2 | 10 | 0.75 | 3.9(3) |  | 64 | 5/1 | 3.1 | 20 | 0.73 | 1.47(5) |
|  | 128 | 22/2 | 2 | 11 | 0.75 | 5.6(6) |  | 128 | 5/1 | 4.4 | 36 | 0.73 | 1.26(5) |
|  | 256 | 22/2 | 2 | 13 | 0.75 | 8.0(7) |  | 256 | 5/1 | 6.2 | 62 | 0.74 | 1.38(9) |
|  | 512 | 22/2 | 2 | 16 | 0.77 | 7.6(6) |  | 512 | 5/1 | 8.8 | 104 | 0.73 | 1.33(9) |
| 1 | 16 | 22/2 | 1.2 | 8 | 0.74 | 3.1(2) | 2.5 | 16 | 10/2 | 1.5 | 8 | 0.82 | 1.46(9) |
|  | 32 | 22/2 | 1.2 | 12 | 0.76 | 3.3(2) |  | 32 | 10/2 | 2.1 | 13 | 0.75 | 1.65(16) |
|  | 64 | 22/2 | 1.2 | 17 | 0.74 | 4.1(2) |  | 64 | 5/1 | 3.0 | 24 | 0.75 | 1.44(16) |
|  | 128 | 36/6 | 1.2 | 24 | 0.75 | 5.5(2) |  | 128 | 5/1 | 4.2 | 42 | 0.77 | 1.19(11) |
|  | 256 | 36/6 | 1.2 | 34 | 0.74 | 6.7(4) |  | 256 | 5/1 | 5.9 | 74 | 0.78 | 1.48(9) |

Table 1: Details of the Monte Carlo runs for a) the unscreened and screened Coulomb models and b) the $r^{-\lambda}$ models with $\lambda = 2$ and 2.5. "Its" indicates, in thousands, the total number of iterations and the number of iterations discarded for thermalization. The statistical error on the acceptance rate was always less than 1%.

## 4 RESULTS

In this section we present the results for the efficiency of the algorithm, which have been obtained using $16 \leq N \leq 512$. We also study the scaling of the end-to-end distance, and compare our results with the predictions mentioned in section 2.

Details of the Monte Carlo runs are given in table 1. The longest runs required about 100 hr CPU time on a DEC 3000. The virial identity, see section 2, was routinely used as a check of the runs. The results have, when possible, been checked against those of refs. [6, 7]. The errors quoted on integrated autocorrelation times have been obtained by dividing the data into eight subsamples. Average and error for the end-to-end distance have been estimated through a jackknife procedure, using between 50 and 200 blocks. We checked that the errors were stable under change in the number of blocks. All errors given are $1\sigma$ errors.

### 4.1 Autocorrelation Times

To monitor the efficiency of the algorithm we have mainly used the integrated autocorrelation time $\tau_{int,ee}$ for the end-to-end distance $r_{1N}$. $\tau_{int,ee}$ gives the cost of generating one independent measurement of $r_{1N}$, assuming that equilibrium has been attained. In several cases, we also studied how the efficiency varied with the length scale considered. This was done by measuring the integrated autocorrelation time $\tau_{int,k}$ for $\sum_\mu \tilde{x}_{k\mu}^2$ for all $k$. The results showed only a fairly weak $k$ dependence, and the maximum $\tau_{int,k}$ was never much larger than $\tau_{int,ee}$. No indication was found of a long autocorrelation time that is missed out when using $\tau_{int,ee}$ to estimate the computational cost. We also note that the $k$ dependence of $\tau_{int,k}$ varies with the simulation parameters. When using shorter trajectory lengths, we observed an increase in $\tau_{int,k}$ at small $k$.

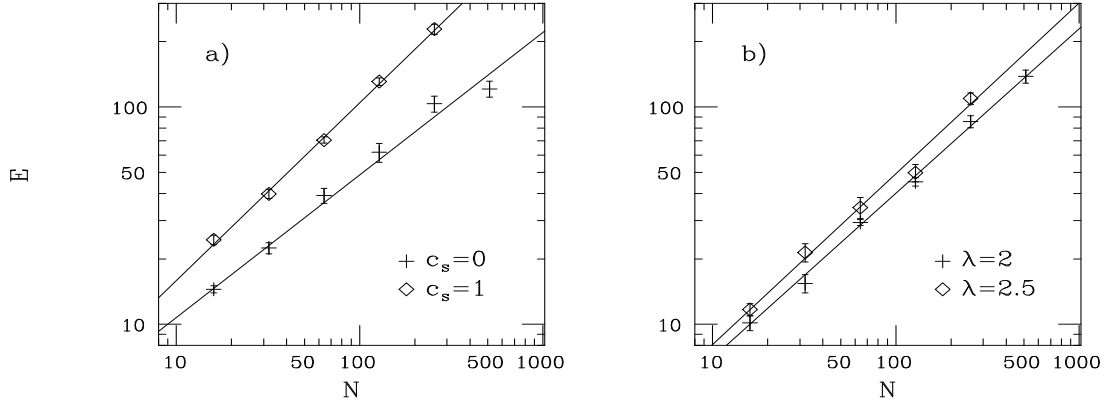

Figure 2: The effort $E$ against $N$ for a) the unscreened and screened Coulomb models and b) the $r^{-\lambda}$ potential with $\lambda = 2$ and 2.5.

| $c_s$ | $\ln c$ | $z'$ | $\chi^2$/d.f. | $\lambda$ | $\ln c$ | $z'$ | $\chi^2$/d.f. |
|---|---|---|---|---|---|---|---|
| 0 | 0.87(9) | 0.66(2) | 2.0 | 2 | 0.21(11) | 0.76(3) | 1.1 |
| 1 | 0.89(10) | 0.82(2) | 0.8 | 2.5 | 0.28(13) | 0.79(3) | 1.8 |

Table 2: Results from fits of the data for $E = n\tau_{int,ee}$ to $E = cN^{z'}$.

In fig. 2, we show the results for $E = n\tau_{int,ee}$. The data are fairly well described by the power law $E \propto N^{z'}$ for all four models. The straight lines in the figure are fits to this form. The details of the fits are given in table 2. There is a clear tendency that the algorithm performs better for the longer range potentials, but the exponent $z'$ does not vary much. The fitted values of $z'$ all lie between 0.64 and 0.84.

## 4.2 End-to-end distance

We begin the study of the scaling with $N$ of the end-to-end distance by forming the effective exponent

$$\nu_N = \frac{1}{2 \ln \frac{N'}{N}} \ln \frac{\langle r_{1\,N'}^2 \rangle_{N'}}{\langle r_{1\,N}^2 \rangle_N} \bigg|_{N'=2N} \tag{20}$$

This gives direct information about the exponents, independently of fitting procedures. If the asymptotic relation in eq. 7 is valid, then we have

$$\nu_N \approx \nu + \alpha \frac{1}{\ln N} \tag{21}$$

for large $N$.

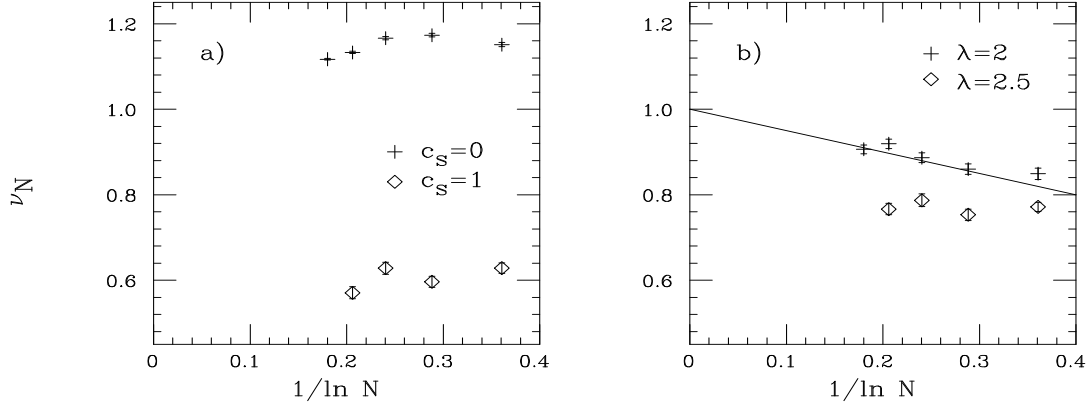

Figure 3: The effective exponent $\nu_N$ for a) the unscreened and screened Coulomb models and b) for the $r^{-\lambda}$ potential with $\lambda = 2$ and 2.5.

Fig. 3 shows our results for $\nu_N$ against $1/\ln N$ in the four different cases. For the screened Coulomb model the results are, although somewhat scattered, close to $\nu = 0.6$. No evidence is seen for corrections to asymptotic pure power-law behaviour. The data for the $r^{-2.5}$ potential are similar. Deviations from pure power-law behaviour are small and the values of $\nu_N$ are close to the predicted value $\nu = 2/\lambda = 0.8$.

The situation is different in the remaining two cases. Here clear deviations from the predicted value of $\nu$ indicate the presence of corrections to the asymptotic pure power law. The straight line shows the predicted large $N$ behaviour for the $r^{-2}$ potential, as given by eq. 7 with $\nu = 1$ and $\alpha = 1/2$. Clearly, the agreement with this prediction is very good, which is confirmed by fits of the data. Using the data for $32 \leq N \leq 512$, a fit to eq. 7 yields $\nu = 1.02(3)$ and $\alpha = 0.58(14)$ with $\chi^2$ per degree of freedom of 1. Because of the large statistical error on $\alpha$, we also performed a fit with $\nu = 1$ fixed, which gave $\alpha = 0.498(12)$ with $\chi^2/\text{d.f.}=0.8$. Hence, the results support the predicted values of both $\nu$ and $\alpha$. For the unscreened Coulomb model, we performed the same types of fits, using again data for $32 \leq N \leq 512$. With $\nu = 1$ fixed we obtained $\alpha = -0.698(3)$ with $\chi^2/\text{d.f.}=3.4$, while the fit of both the exponents gave $\nu = 0.997(7)$ and $\alpha = -0.71(4)$ with $\chi^2/\text{d.f.}=4.9$. The quality of these fits is not perfect. The reason could be that the asymptotic form is still a bad approximation at $N = 32$, as in fact suggested by the figure. When restricting the one-exponent fit to $128 \leq N \leq 512$, we obtained $\alpha = -0.689(6)$ and indeed a very small $\chi^2/\text{d.f.}$. The data therefore seem completely consistent with eq. 7 also for the unscreened Coulomb model, and suggest that $\alpha \approx -0.7$. The negative sign of $\alpha$ implies that the average distance between neighbouring monomers diverges as $N \to \infty$.

## 5 SUMMARY

We have developed the HMC method for simulation of single off-lattice chains with interactions between all monomer pairs. The method is exact and makes it possible to update all the degrees of freedom in a computer time of order $N^2$. We have tested the performance of the algorithm on models with short- or long-range self-repulsion. We found that these models can be simulated in a computer time of order $N^{2+z'}$ with $z'$ between 0.64 and 0.84. These estimates are for measurements of local as well as global quantities. The fact that the efficiency is similar on different length scales distinguishes HMC from currently used algorithms. This property makes it easier to control thermalization, which we for large $N$ expect to be faster with HMC. Possible ways to further improve the efficiency of the algorithm include a systematic fine tuning of the trajectory length and the use of a higher-order discretization scheme [17, 18, 19].

We have in this paper restricted our attention to linear chains with free endpoints. We have discussed two versions of HMC which are equivalent for such chains, namely the bond variable formulation and the Fourier accelerated algorithm. The applicability to more general topologies is different for these two methods. Bond variables can be directly applied to branched structures, while the Fourier acceleration technique is well suited for the study of ring polymers.

We have also presented results for the scaling of the end-to-end distance. For two of the models studied, we found evidence for corrections to asymptotic pure power-law behaviour. Our results for the $r^{-2}$ potential support the asymptotic relation $\langle r_{1N} \rangle \sim N^\nu (\ln N)^{-\alpha}$ with $\nu = 1$ and $\alpha = 1/2$, which was predicted in ref. [8]. The results for the unscreened Coulomb model are well described by the same expression with $\nu = 1$ and $\alpha \approx -0.7$.

**Acknowledgments**

I would like to thank Bo Jönsson, Carsten Peterson and Bo Söderberg for useful discussions.